# On the elicitation of continuous, symmetric, unimodal distributions

John Paul Gosling

May 30, 2018

## Abstract

In this brief note, we highlight some difficulties that can arise when fitting a continuous, symmetric, unimodal distribution to a set of expert's judgements. A simple analysis shows it is possible to fit a Cauchy distribution to an expert's beliefs when their beliefs actually follow a normal distribution. This example stresses the need for careful distribution fitting and for feedback to the expert about what the fitted distribution implies about their beliefs.

*Keywords: Distribution fitting, expert elicitation, heavy-tailed distribution, prior misspecification.*

## 1 Elicitation

In the context of statistical analysis, *elicitation* is the process of translating someone's beliefs about some uncertain quantities into a probability distribution. An elicited probability distribution is often used as a prior distribution in a Bayesian analysis. In some application areas, there are few or no data available and expert opinion becomes paramount in the decision making process. See Ang and Buttery (1997), Coolen et al. (1992) and Chaloner and Rhame (2001) for examples in nuclear safety, engineering and clinical trials respectively.

Elicitation is far from being a precise science. It can be difficult for experts to articulate their beliefs. There are also other complications due to the possible biases of the experts and the biases created by the questioning process. The process of questioning people about their beliefs is certainly not a new subject: it has been the focus of many psychological studies. Reviews of the psychological literature from a statistical viewpoint can be found in Kadane and Wolfson (1998) and O'Hagan et al. (2006). In this note, we use a simple example to show the implications to prior modelling of an expert's inability to articulate their beliefs with absolute precision. Throughout this note, we assume that the expert being questioned is female for ease of exposition and that there is a facilitator of the elicitation process conducting the elicitation exercise.

[1]Department of Probability and Statistics, University of Sheffield, Sheffield, UK, j.p.gosling@sheffield.ac.uk

## 2 Fitting distributions to percentile judgements

The expert's beliefs for some continuous quantity $X$ are to be modelled. Her density function for $X$ is then $f(X)$. The facilitator elicits information from her about $f(.)$. She is not expected to accurately report the value of $f(\theta)$ for all possible values of $\theta$. In fact, she should not be expected to be able to report $f(\theta)$ for any value of $\theta$. Kadane and Wolfson (1998) suggest that quantiles or probabilities should be elicited, and, often, the expert is comfortable only providing the median and quartiles.

For $X$, the expert is able to give the facilitator the following judgements: $P(X < -0.6745) = 0.25$, $P(X < 0) = 0.5$ and $P(X < 0.6745) = 0.75$; these are obviously the median and quartiles of the standard normal distribution. In addition to this, suppose that she is happy for the facilitator to assume that her density for $X$ is unimodal. From the quartile judgements, a symmetric distribution would be an appropriate fit to her judgements. There are infinitely many distributions that would satisfy these conditions. In Figure 1, three possible cumulative distribution functions that interpolate her judgements precisely are plotted. The three distributions have widely varying tail-behaviours .

In this example, we could try to discriminate between the possible fits by eliciting more judgements from the expert. Several methods have been proposed to get information about the tails of an expert's distribution: tail ratios are discussed in Kadane et al. (1980) and hypothetical data are discussed in Garthwaite and Al-Awadhi (2001). In these methods, further judgements are required from the expert in order to discover more about the tails of the distribution: in Kadane et al. (1980), a judgement about the $90^{\text{th}}$ percentile are used to help characterise tail-behaviour. In our example, we suppose that the expert is able to specify $P(X < -1.2816) = 0.1$ and $P(X < 1.2816) = 0.9$ in addition to her earlier judgements.

If we are uncertain as to whether the reported $i^{\text{th}}$ percentile is in fact the expert's $(i-2.5)^{\text{th}}$ percentile or the $(i+2.5)^{\text{th}}$ percentile, then we should take account of this uncertainty. In fact, in many cases, the expert is only expected to be able to make probability judgements up to a precision of 0.05 (see O'Hagan et al., 2006, for examples). In addition to this, we may suspect that she may only be able to give judgements about $X$ to within $\pm 0.05$. This seems reasonable as this creates an interval that is a tenth of the her 'true' distribution's standard deviation.

The graph in Figure 2 shows the results of fitting distributions to the expert's judgements where independent uniform distributions are used to represent our uncertainty about the judgements. It is clear that we still could fit practically any unimodal symmetric distribution to her judgements. The true situation may be worse than what is depicted here as it is reasonable to expect the uncertainty to grow as she makes judgements further into the tails of her distribution.

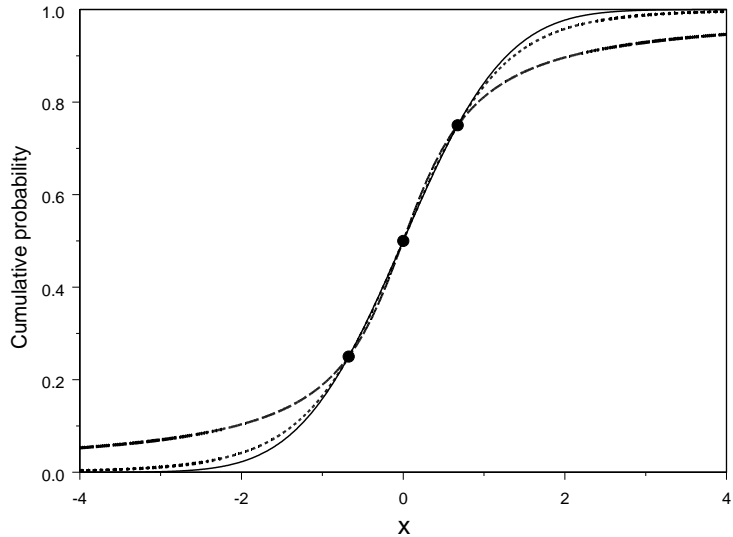

Figure 1: The dots are the expert's judgements and the three lines are possible fits: normal (solid), $t$ with 5 degrees of freedom (dotted) and Cauchy (dashed).

## 3 Summary

Similar fitting procedures using percentile judgements are common in Bayesian analyses; normal distributions were fitted to this type of expert judgement in Cooke and Slijkhuis (2003), Denham and Mengersen (2007) and Kennedy et al. (2008). It is clear from the example of the previous section that the facilitator should check the sensitivity of their results to the possible tail-behaviours. Often, we suspect that the tail behaviour will make little difference; however, if there are no data or the data are far from what is expected *a priori*, then the tail-behaviour in the elicited prior could have serious implications (see Andrade and O'Hagan, 2006).

The models for elicitation described in O'Hagan and Oakley (2004) and Gosling et al. (2007) begin to address some of the uncertainties in the elicitation process; more specifically, the facilitator's uncertainty in the choice of distribution to fit. How-ever, uncertainty about the precision of the expert's judgements is not given full attention. One way to help prevent misrepresentation of an expert's beliefs is to include feedback stages in the elicitation exercise. During the feedback stage, the expert is shown the implications of using a particular distribution (for example, the tertiles of the fitted distribution) and asked if they wish to revise her original judgements and/or make additional judgements. Unfortunately, this additional procedure cannot eliminate the expert's inability to specify her judgement's with absolute precision. In conclusion, we reiterate that elicitation is not a precise science, and, in order to make progress, more effort should be made to develop elicitation models that accommodate all the forms of uncertainty that are clearly present in elicitation.

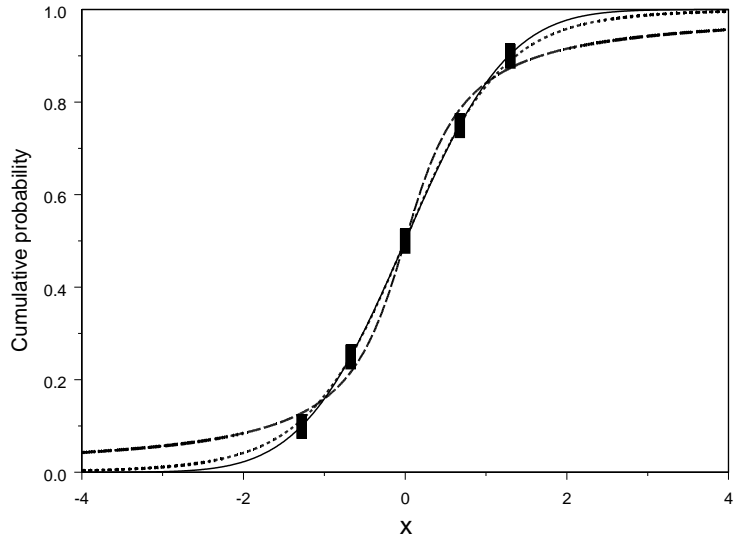

Figure 2: The rectangles represent uniform uncertainty about the expert's five judgements and the three lines are possible fits: normal (solid), $t$ with 5 degrees of freedom (dotted) and Cauchy (dashed).

# References

J.A.A. Andrade and A. O'Hagan. Bayesian robustness modelling using regularly varying distributions. *Bayesian Analysis*, 1:169–188, 2006.

M.L. Ang and N.E. Buttery. An approach to the application of subjective probabilities in level 2 psas. *Reliability Engineering and System Safety*, 58:145–156, 1997.

K. Chaloner and F.S. Rhame. Quantifying and documenting prior beliefs in clinical trials. *Statistics in Medicine*, 20:581–600, 2001.

R.M. Cooke and K.A. Slijkhuis. Expert judgment in the uncertainty analysis of dike ring failure frequency. In *Case Studies in Reliability and Maintenance* (eds. W.R. Blischke and D.N. Prabhakar Murthy), pages 331–352. Chichester: Wiley, 2003.

F.P.A Coolen, P.R. Mertens, and M.J. Newby. A bayes-competing risk model for the use of expert judgement in reliability estimation. *Reliability Engineering and System Safety*, 35:23–30, 1992.

Robert Denham and Kerrie Mengersen. Geographically assisted elicitation of expert opinion for regression models. *Bayesian Analysis*, 2:99–136, 2007.

Paul H. Garthwaite and Shafeeqah A. Al-Awadhi. Non-conjugate prior distribution assessment for multivariate normal sampling. *J. R. Statist. Soc. Ser. B*, 63:95–110, 2001.

John Paul Gosling, Jeremy E. Oakley, and Anthony O'Hagan. Nonparametric elicitation for heavy-tailed prior distributions. *Bayesian Analysis*, 2:693–718, 2007.

J.B. Kadane and L.J. Wolfson. Experiences in elicitation. *The Statistician*, 47:3–19, 1998.

Joseph B. Kadane, James M. Dickey, Robert L. Winkler, Wayne S. Smith, and Stephen C. Peters. Interactive elicitation of opinion for a normal linear model. *Journal of the American Statistical Association*, 75(372):845–854, 1980. ISSN 01621459.

Marc Kennedy, Clive Anderson, Anthony O'Hagan, Mark Lomas, Ian Woodward, John Paul Gosling, and Andreas Heinemeyer. Quantifying uncertainty in the biospheric carbon flux for england and wales. *Journal of the Royal Statistical Society: Series A (Statistics in Society)*, 171:109–135, 2008.

A. O'Hagan, C.E. Buck, A. Daneshkhah, J.E. Eiser, P.H. Garthwaite, D. Jenkinson, J.E. Oakley, and T. Rakow. *Uncertain judgements: eliciting expert probabilities*. Chichester: Wiley, 2006.

Anthony O'Hagan and Jeremy E. Oakley. Probability is perfect, but we can't elicit it perfectly. *Reliability Enigineering and System Safety*, 85:239–248, 2004.